\documentclass[11pt]{article}
\usepackage[T1]{fontenc}
\usepackage[utf8]{inputenc}
\usepackage{lmodern}
\usepackage{amsmath}
\usepackage{amssymb}
\usepackage{graphicx}
\usepackage{float}
\usepackage{geometry}
\usepackage[numbers]{natbib}
\usepackage[colorlinks=true, linkcolor=blue, urlcolor=blue, citecolor=blue]{hyperref}
\title{Deep brain microelectrode signal: $q$-statistical approach}
\author{%
Ana Luiza Souza Tavares$^{1}$, Henrique Santos Lima$^{2}$, Artur Pedro Martins Neto$^{1}$,\\
Bruno Duarte Gomes$^{1,3,4}$, Constantino Tsallis$^{2,3,4,5}$\\
\small $^{1}$Instituto de Ci\^encias Biol'ogicas, Universidade Federal do Par'a, Bel'em, Par'a, Brazil\\
\small $^{2}$Centro Brasileiro de Pesquisas F'isicas, Rio de Janeiro, Brazil\\
\small $^{3}$Santa Fe Institute, Santa Fe, New Mexico, USA\\
\small $^{4}$Complexity Science Hub Vienna, Vienna, Austria\\
\small $^{5}$Dipartimento di Fisica e Astronomia Ettore Majorana, Catania, Sicily, Italy\\
\small \texttt{ana.souza.tavares@icb.ufpa.br}, \texttt{hslima94@cbpf.br}, \texttt{artur.neto@icb.ufpa.br},\\
\small \texttt{brunodgomes@ufpa.br}, \texttt{tsallis@cbpf.br}}
\date{}

\begin{document}
\maketitle

\begin{abstract}
 We characterize the amplitude statistics of intraoperative microelectrode recordings (MERs) obtained during deep brain stimulation (DBS) 
surgery in 46 patients with Parkinson's disease, using 184 recordings equally 
balanced between inside and outside the subthalamic nucleus (STN). The probability 
density of every recording is quantitatively well described by a $q$-Gaussian 
(grounded on a nonadditive entropic functional), $\rho(x) \propto [1 + \beta(q-1) 
x^2]^{-1/(q-1)}$, with $q > 1$ in all cases, reflecting persistent long-range 
temporal correlations inconsistent with Gaussian dynamics. Within the superstatistics 
framework, the slowly fluctuating local variance visible in the raw MER signals is 
a physical mechanism that directly generates the $q > 1$ form. Beyond individual 
fits, $q$ and $\beta$ collapse across all 184 recordings onto the single functional 
constraint $q = 3 - 1.85\,\beta^{-0.33}$ ($R \approx -0.91$), a reduction to 
one effective degree of freedom that is the quantitative hallmark of near-critical 
dynamics, previously identified in scale-free network growth and in acoustic 
precursors of material fracture. The index $q$ is statistically indistinguishable across the STN boundary ($\langle\bar{q}_\text{out}/\bar{q}_\text{in} \rangle = 1.03$), while the inverse-widthparameter shows a modest systematic difference ($\langle\bar{\beta}_\text{out}/\bar{\beta}_\text{in} \rangle = 1.18$). Since $q > 1$ is expected for any brain structure exhibiting long-range correlations, healthy or pathological, it is the tight $q(\beta)$ coupling, not $q > 1$ per 
se, that constitutes the candidate near-criticality signature of the parkinsonian 
cortico-basal-ganglia-thalamocortical loop.
\end{abstract}


\section{Introduction}

Deep Brain Stimulation (DBS) of the subthalamic nucleus (STN) has become an established surgical treatment for advanced Parkinson's disease (PD), providing substantial improvement of motor symptoms and reduction of medication requirements through the delivery of high-frequency electrical stimulation to the cortico-basal-ganglia-thalamocortical loop \cite{hutchison, beudel,eusebio}. In the healthy state, this circuit operates as a selective filter and gain-control device that amplifies intended motor commands while suppressing competing ones; in the parkinsonian condition, dopamine depletion along the indirect pathway disrupts this balance, giving rise to pathological, synchronized beta-band oscillations in the STN that are tightly correlated with bradykinesia and rigidity \cite{chiken}. A critical step in DBS surgery is the accurate localization of the STN, and intraoperative microelectrode recordings (MERs) are widely employed to refine targeting by monitoring characteristic changes in neuronal firing patterns as the electrode traverses surrounding structures \cite{vissani}. These signals provide neurophysiological landmarks that complement imaging, but their interpretation remains challenging due to stochastic variability and non-stationary dynamics.

Conventional MER analysis typically relies on descriptive features such as spike rates, firing regularity, or spectral energy. While clinically useful, these measures are insensitive to the statistical geometry of the amplitude distribution. In particular, they cannot distinguish between distributions with light tails, consistent with weakly correlated Gaussian dynamics, and heavy-tailed distributions that arise when strong, non-local correlations are present \cite{chialvo, beggs}. Recent studies have highlighted the importance of entropy-based measures to quantify neural signal complexity; however, most existing frameworks employ Shannon-type entropies, which rest on the assumption of additivity and short-range correlations \cite{keshmiri, luczak}. This assumption is in general unlikely to hold in the STN and surrounding basal ganglia structures, given that the cortico-basal-ganglia-thalamocortical network operates as a dynamically coupled system with persistent, long-range temporal correlations, precisely the statistical regime in which the Boltzmann-Gibbs–Shannon framework loses its domain of validity. In the parkinsonian condition, dopamine depletion along the indirect pathway further constrains this network into a pathologically synchronized dynamical regime, most visibly expressed as stereotyped beta-band bursting, which makes the departure from additivity not only expected but measurable with intraoperative MER signals \cite{chiken}.

The nonadditive entropic functional  $S_q=k_B(1-\sum_{i} p_i^q)/(q-1)$ extends the Boltzmann–Gibbs–von Neumann-Shannon formalism to systems with long-range interactions, memory effects, or fractal phase-space structures. Within this framework, stationary states are described by q-statistics, and probability distributions of standard fluctuations take the $q$-Gaussian form (Eq. (1)), which, for $q>1$ and $q<3$, defines a normalizable, heavy-tailed distribution that decays asymptotically as a power law $\rho(x)\propto x^{-2/(q-1)}$ \cite{Tsallis1988, tsallis2023}. The index $q$ is not merely a shape parameter: it quantifies the degree of departure from Gaussian statistics and, by extension, the relevance of long-range correlations in the system's dynamics. Within the $q$-statistical formalism, the full nonequilibrium characterization of a system requires the determination of the so-called q-triplet $(q_{sen},q_{rel},q_{stat})$, whose three indices govern, respectively, the extensivity of $S_q$, the relaxation of temporal correlations, and the stationary probability distribution. The present work focuses on $q_{\mathrm{stat}}$, estimated directly from the amplitude distributions of the MER signals, leaving the reconstruction of the full triplet as a natural extension.
\section{Model and Methods}
\subsection{Dataset}
The data used is an open-access intraoperative MER dataset collected during DBS procedures for Parkinson’s disease, published by Ciecierski and colleagues \cite{dataset}. The dataset includes 46 patients from a single center, with 4,650 recordings of approximately 10\,s each; curator labels indicate 3,210 recordings outside the STN and 1,440 inside the STN. For this exploratory study, we used a patient-balanced subsample: four recordings per patient, two labeled inside STN and two outside STN, for a total of 184 recordings. For every combination of patient and label, multiple candidate recordings were available, and we randomly selected two recordings from these candidates.

\subsection{Preprocessing pipeline}
All MERs were pre-processed using a standardized pipeline to remove non-physiological segments and preserve the quality of neuronal activity. 

The first step consisted of an edge-trimming procedure based on the analytic envelope and temporal derivative of each signal. The envelope was computed via the Hilbert transform, and its median and median absolute deviation (MAD) were estimated over the central 60\% of the recording. The median envelope and derivative values in the initial and final 300\,ms were compared to these central statistics. Segments displaying envelope values greater than the upper threshold $\mathrm{Med}_e\,+\,6\,\mathrm{MAD}_e$, lower than the lower threshold $\mathrm{Med}_e\,-\,2.5\,\mathrm{MAD}_e$, or sustained derivative values below a flatline threshold $|x'(t)| < 0.08\,\mathrm{MAD}_e $ were removed, respectively capturing abrupt high-amplitude transients, abnormally low-energy silent or padded regions, and prolonged periods of minimal temporal variation indicative of acquisition-induced plateaus. Trimming proceeded until both metrics returned to within normal ranges, with a minimum retention of 2\,s to avoid excessive exclusion.

The trimmed signal was then band-pass filtered between 300 and 5000 Hz using a fourth-order Butterworth filter applied in zero-phase mode, preserving MER activity while suppressing low-frequency drift and high-frequency noise.

Residual artifacts were addressed with an adaptive envelope-based correction. The histogram of the analytic envelope was used to estimate a robust noise standard deviation by fitting a quadratic model to the region surrounding the histogram peak. Samples whose envelopes exceeded eight times this estimate were classified as artifacts and replaced using monotonic cubic interpolation (PCHIP) across neighboring clean samples, preserving continuity without discarding data.

Finally, each cleaned signal was z-score normalized to remove amplitude variability across recordings.

\subsection{Fitting procedure}

We model the processed  inside and outside distributions with a $q$-Gaussian written in the normalized form:
\begin{equation}
\frac{p(x)}{p_0}=e_q^{-\beta x^2},
\end{equation}
where $p_0\equiv p(0)$ is the maximum of $p$, $\beta$ is an inverse-width parameter, and the $q$-exponential function is defined as $e_q^{z}\equiv [1+(1-q)z]^{1/(1-q)}\;\;(e_1^z=e^z)$.

Introducing its inverse function, namely the $q$-logarithm,
\begin{equation}
\ln_q(y)=\frac{y^{1-q}-1}{1-q},
\qquad
\ln_1(y)=\ln(y),
\end{equation}
the model can be represented as follows:
\begin{equation}
\ln_q\!\left[\frac{p(x)}{p_0}\right]=-\beta x^2.
\end{equation}
Therefore, for the optimal choice of parameters, the transformed data should fall approximately on a straight line in the $(x^2,\ln_q(p/p_0))$ representation.

In practice, for each dataset we compute the empirical ratio $p(x)/p_0$ and build the transformed variables $X=x^2$ and $Y=\ln_q(p/p_0)$. The parameters $(q,\beta)$ are then determined via a grid-search procedure: we scan a predefined mesh of trial values $(q,\beta)$ and quantify the degree of linearity of the corresponding $(X,Y)$ scatter plot.

For each candidate pair $(q,\beta)$, we perform a linear regression of $Y$ versus $X$ and compute the coefficient of determination $R^2$. The optimal parameters $(q^\star,\beta^\star)$ are defined as those maximizing $R^2$, i.e., the pair that produces the closest-to-linear scaling of $\ln_q(p/p_0)$ as a function of $x^2$ within the explored parameter grid.

After determining each pair $(q,\beta)$, we then perform a linear regression using the variables $(q,1/\beta^{\alpha})$. The aim is to determine the value of the exponent $\alpha$ that produces the best straight line for the set of $(q,1/\beta^{\alpha})$ values, again using the $R^2$ criterion.


\section{Results}
\subsection{Data \& Preprocessing Overview}
Across all 46 patients, 184 MER recordings were processed by the pipeline removing non-physiological artifacts. Most recordings preserved their full duration, with 54.4\% retaining all samples, with the shortest valid recording exceeding 3.3 seconds. Envelope-based artifact detection identified low contamination overall, with a mean of 1.74\% $\pm$ 2.07\% of samples interpolated and a median of 1.02\%. These results demonstrate that the preprocessing pipeline removed boundary artifacts and transient disturbances only when necessary, while preserving the physiological integrity and temporal extent of the MER signals.

\begin{figure}[h]
    \centering
    \includegraphics[width=16cm]{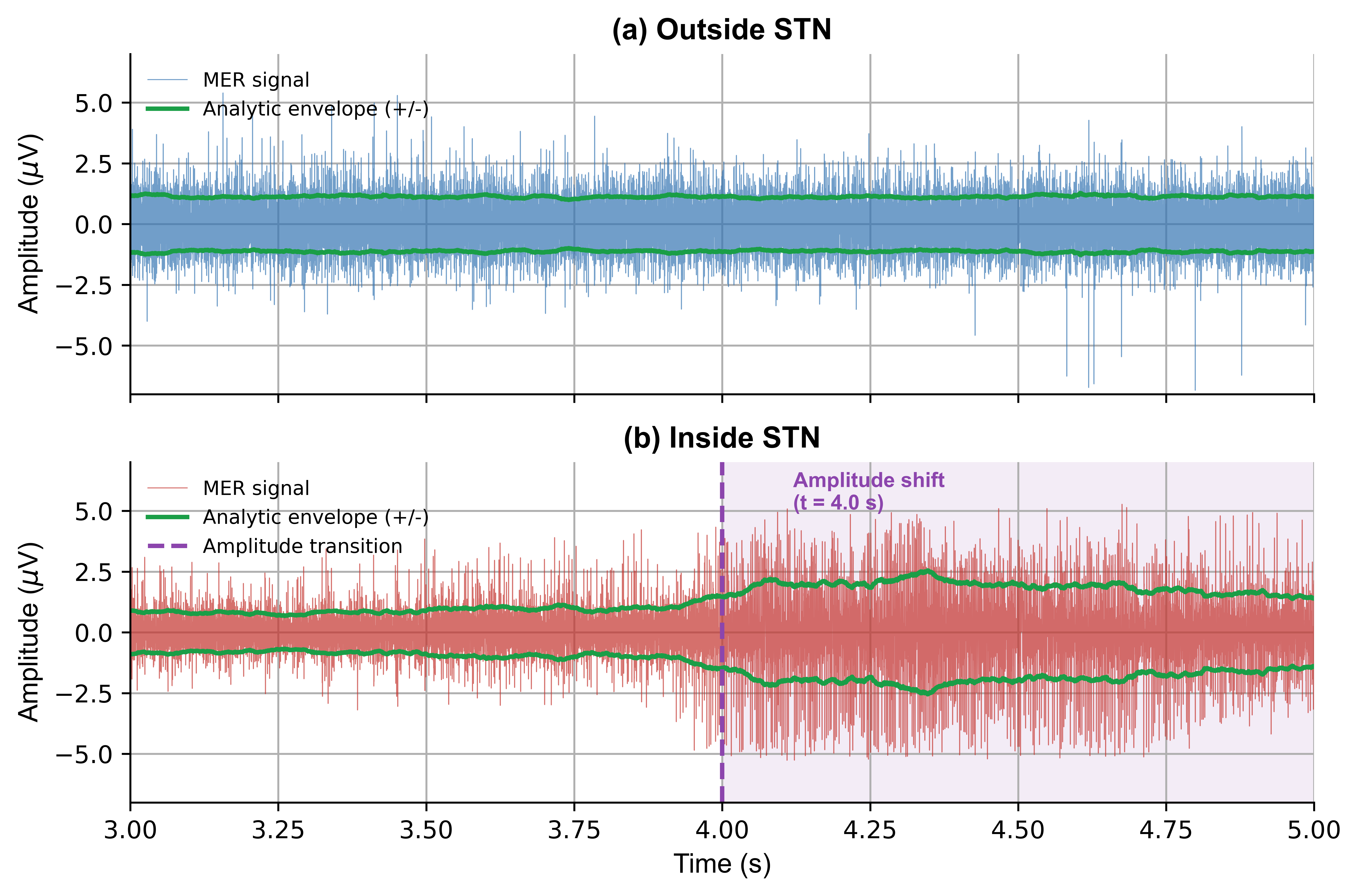}
    \caption{Raw microelectrode recording amplitudes ($\mu$V) as a function of recording time (s) for a representative Parkinson's disease patient, following bandpass filtering (300--5000~Hz) and $z$-score normalisation. (a)~Extra-STN recording; (b)~intra-STN recording. The sustained amplitude elevation visible near the midpoint of panel~(b) reflects the within-recording non-stationarity discussed in the text, consistent with slow variance fluctuations that physically generate the $q > 1$ form within the superstatistics framework. }
    \label{fig1}
\end{figure}

\subsection{Evidence of near-critical nonextensive dynamics}
\begin{figure}[H]
    \centering
    \includegraphics[width=7.5cm]{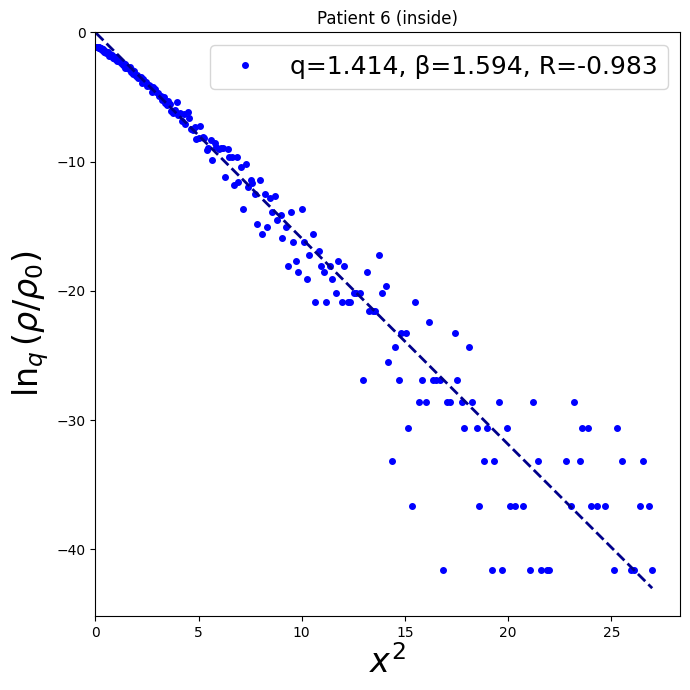}
    \includegraphics[width=7.5cm]{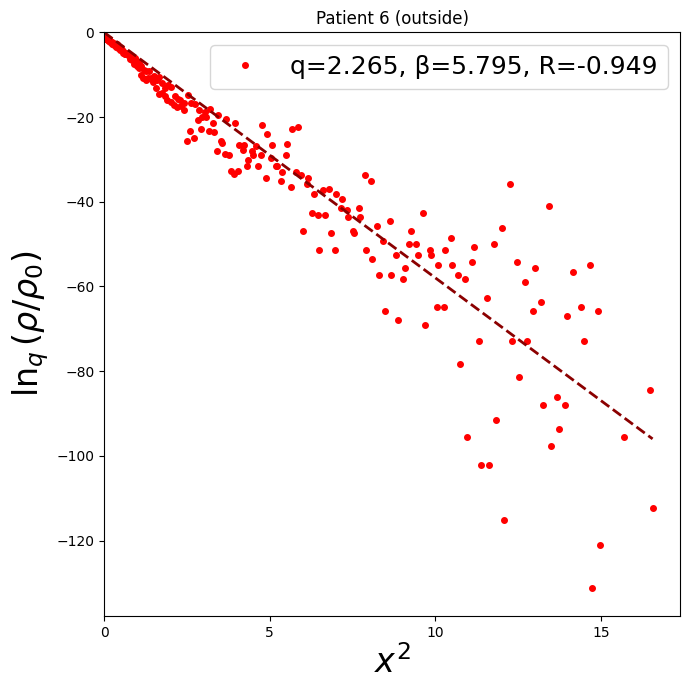}
    \includegraphics[width=7.5cm]{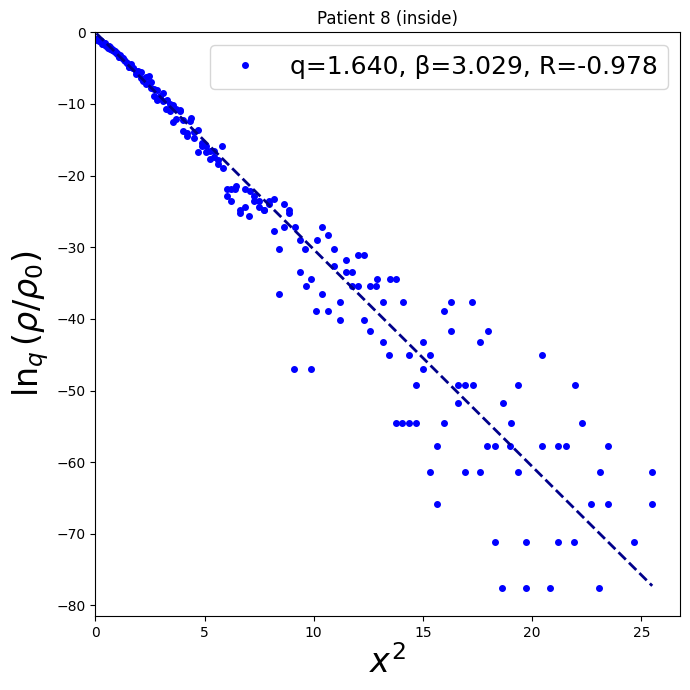}
    \includegraphics[width=7.5cm]{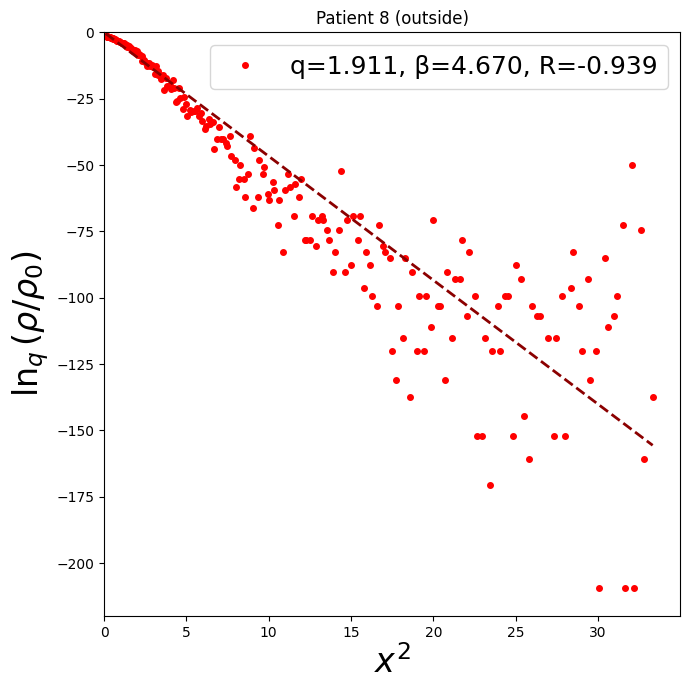}
    \caption{$q$-Gaussian fitting of MER amplitude distributions for two representative Parkinson's disease patients. The $q$-logarithmic transform $\ln_q[\rho(x)/\rho_0]$ is plotted against the squared normalised amplitude $x^2$, so that a $q$-Gaussian
    $\rho(x) = \rho_0\, e_q^{-\beta x^2}$ appears as a straight line with slope $-\beta$ (dashed lines). (a)~Patient~4;
    (b)~Patient~6. Blue symbols: intra-STN recordings; red symbols: extra-STN recordings. Fitted parameters $(q^{\star}, \beta^{\star})$ and the coefficient of determination $R^2$ for each fit are annotated within the panels..}
    \label{fig2}
\end{figure}
\begin{figure}[h]
    \centering
    \includegraphics[width=12cm]{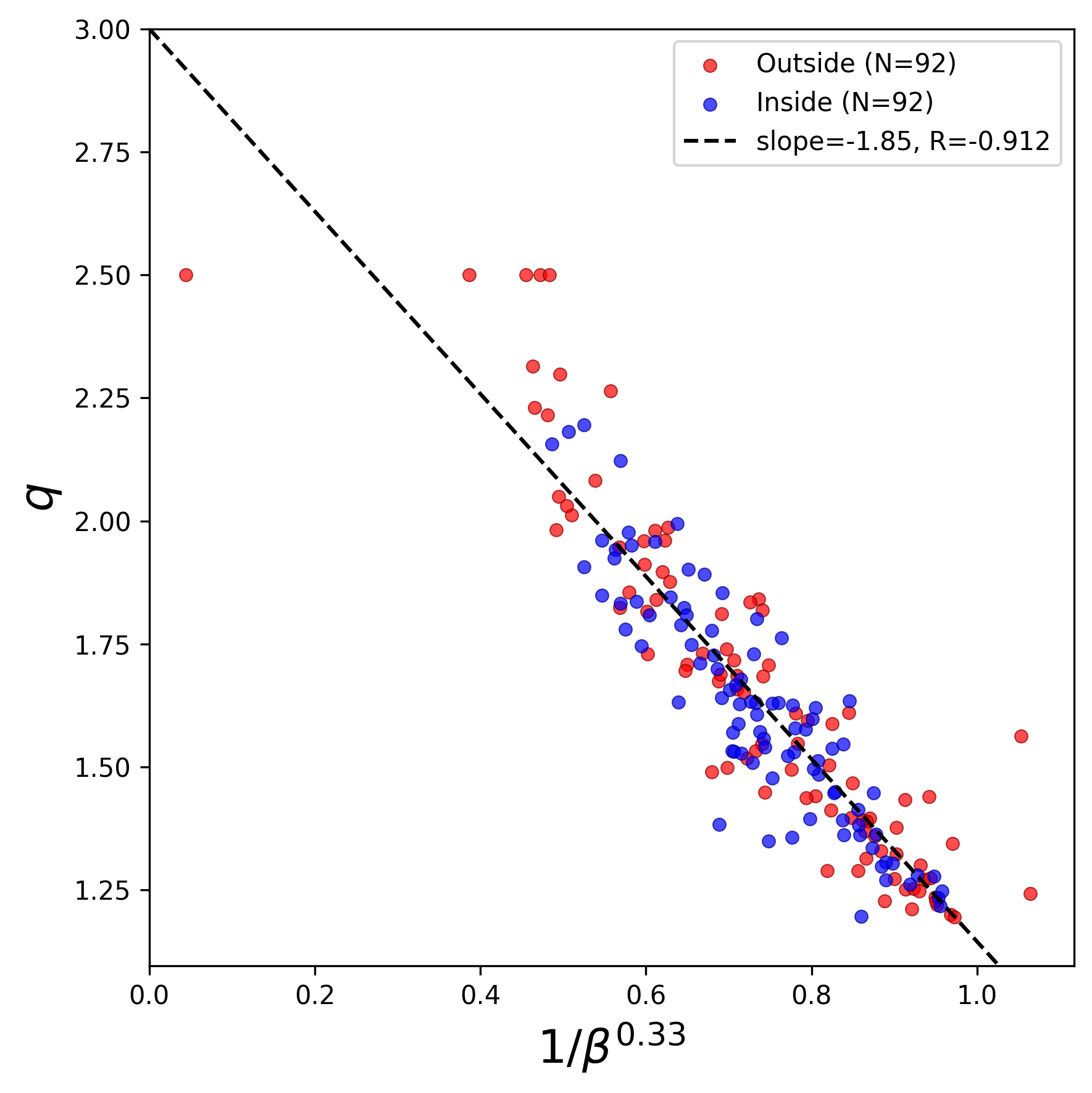}
    \caption{$q$ versus $1/\beta^{0.33}$ for 184 signals. The dashed straight line  (black) is the linear fitting $(q=3-1.85/\beta^{0.33})$ of the outside (red circles) and inside (blue squares) datasets  assuming $q$-Gaussian forms (notice that $q$-Gaussians are normalizable only if $q < 3$). The  linear correlation coefficient $R \simeq - 0.912$ indicates a strong correlation between  the values of $q$ and $\beta$. A monotonic behavior like this one characterizes criticality and has already been observed in the growth of asymptotically scale-free networks\cite{BritoSilvaTsallis2016,NunesBritoSilvaTsallis2017,CinardiRapisardaTsallis2020} and in the fracture of construction materials\cite{GrecoTsallisRapisardaPluchinoFicheraContrafatto2020}.}
    \label{fig3}
\end{figure}

Figure~\ref{fig1} illustrates the raw MER signals for an arbitrarily selected patient, showing the full 10-second recording both outside (panel a) and inside (panel b) the STN. A notable feature of the inside-STN signal is a sustained increase in amplitude occurring around the mid-point of the recording, after which the signal maintains an elevated amplitude envelope. This behavior is consistent with a well-documented source of non-stationarity in intraoperative MERs, a progressive loss of cerebrospinal fluid following dural opening gives rise to a slow downward brain shift that alters the spatial relationship between the fixed electrode tip and nearby neuronal populations \cite{petersen2010,miyagi2007}, while superimposed cardiac pulsatility modulates the electrode-tissue interface continuously throughout the recording \cite{lempka2009}, together producing step-like or gradually sustained changes in recorded amplitude. An intrinsic neurophysiological contribution is equally plausible. STN neurons in the dopamine-depleted state are well known to transition between tonic and sustained bursting regimes within individual recording epochs \cite{hutchison}, and burst features are preserved even under general anesthesia \cite{lin2017}, so that the onset of a prolonged bursting episode would manifest as precisely the kind of sustained amplitude elevation observed in Fig. 1b. Importantly, this within-recording amplitude non-stationarity is not merely a limitation to be acknowledged \cite{vissani}, but is theoretically germane to the $q$-statistical results that follow. Within the superstatistics framework of Beck and Cohen \cite{beck2003}, a process whose local variance fluctuates slowly in time as is visually manifest in Fig. 1b yields a marginal amplitude distribution that is q-Gaussian with $q > 1$. The non-stationarity of the MER signal is therefore not in tension with the $q$-Gaussian description; it is one of its physical mechanisms.

In Fig.~\ref{fig2}, we present the results for $\ln_q (\rho / \rho_0)$ versus $x^2$ for patients 4 and 6, both in the inside and outside regions. The datasets align closely with the fitted straight lines, as evidenced by $R^2$ values near 1. These findings confirm that the probability distributions decay asymptotically as a power law,
\begin{equation}
\rho(x) \propto x^{-2/(q-1)},
\end{equation}
with a $\beta$-dependent constant of proportionality.

The parameters $q$ and $1/\beta^{0.33}$, obtained from the inside and outside fittings, nearly collapse onto a single  (monotonic) curve (far from a disperse cloud) adequately fitted by a straight line with $R\approx-0.91$ (Fig.~\ref{fig3}). This collapse reveals a functional dependence $q = q(\beta)$, explicitly given by
\begin{equation}
q(\beta) = 3 - \frac{1.85}{\beta^{0.33}},
\end{equation}
indicating that $q$ and $\beta$ are not independent. Consequently, the universal exponent
\begin{equation}
\frac{2}{q(\beta)-1} = \frac{2}{2 - 1.85 \, \beta^{-0.33}}
\end{equation}
diverges for decreasing $\beta$ (thus approaching a Gaussian behavior) and tends to 1 as $\beta \to \infty$, recovering a Zipf-like decay $\rho \propto x^{-1}$. For finite values of $\beta$, the distributions invariably exhibit asymptotic power-law tails. The fact that, for a given value of $q$, we nearly have an unique value of $\beta$ is of paramount significance. Indeed, it means that the system is nearly criticality. This functional dependence is a hallmark of criticality, previously documented in asymptotically scale-free networks and in material fracture.

\subsection{Spatial invariance of nonextensive dynamics across the STN boundary}

\begin{figure}[htb]
    \centering
    \includegraphics[width=14cm]{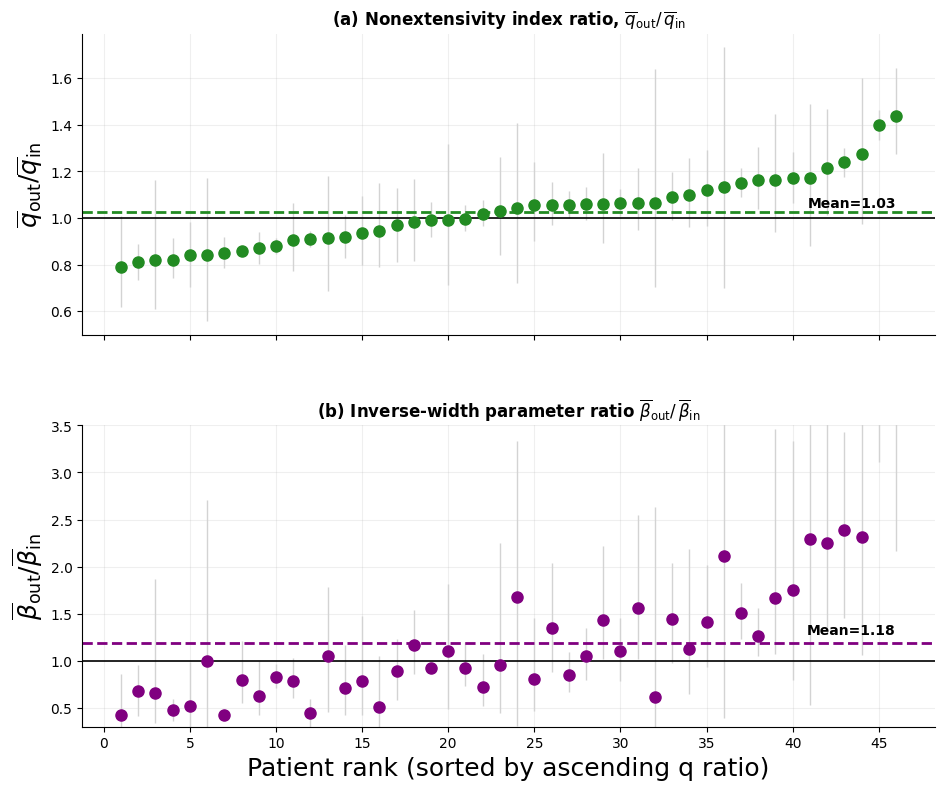}
    \caption{Patient-wise ratios of mean fitted $q$-Gaussian parameters between extra-STN and intra-STN recordings. For each of the 46 patients ($x$-axis: patient index), green circles show the ratio $\bar{q}_{\text{out}}/\bar{q}_{\text{in}}$ of patient-averaged nonextensivity indices, and purple circles show the ratio
    $\bar{\beta}_{\text{out}}/\bar{\beta}_{\text{in}}$ of
    patient-averaged inverse-width parameters. The dashed horizontal line marks unity. Ensemble means are $\langle \bar{q}_{\text{out}}/\bar{q}_{\text{in}} \rangle = 1.03$ and $\langle \bar{\beta}_{\text{out}}/\bar{\beta}_{\text{in}} \rangle = 1.18$, confirming spatial uniformity of the nonextensivity index $q$ across the STN boundary, while the modest elevation of $\bar{\beta}_{\text{in}}$ reflects the well-documented
    increase in background spiking amplitude within the STN.}
    \label{fig4}
\end{figure}

Fig.~\ref{fig4} presents, for each of the 46 patients, the ratios of the ensemble-averaged fitted parameters between recordings acquired outside and inside the STN. The green markers display $\bar{q}_{outside}/\bar{q}_{inside}$ and scatter tightly around unity, yielding an ensemble mean of $\left\langle \bar{q}_{outside}/\bar{q}_{inside} \right\rangle=1.03$. This near-perfect ratio carries a clear and striking message. The nonextensivity index $q$, that encodes the degree of departure from Gaussian statistics and, by extension, the intensity of long-range correlations in the signal, is statistically indistinguishable between recordings acquired within and outside the target nucleus. The purple markers, tracking $\bar{\beta}_{outside}/\bar{\beta}_{inside}$, tell a slightly different story: their ensemble mean of $\left\langle \bar{\beta}_{outside}/\bar{\beta}_{inside} \right\rangle =1.18$ indicates that outside-STN signals carry modestly narrower amplitude distributions — that is, their generalized variance $1/\beta$ is somewhat smaller than that observed within the STN. This modest elevation of the intrinsic signal energy inside the nucleus is consistent with the well-documented increase in background spiking activity and local field fluctuation amplitude that characterizes the MER signature of the STN proper. What is remarkable, however, is that this energy-level difference occurs in the complete absence of any commensurate change in $q$. The statistical character of the fluctuations, their heavy-tailed, power-law structure, their degree of non-Gaussianity, remains essentially constant across the STN boundary. Taken together, the results from Figs. 3 and 4 establish two complementary and mutually reinforcing conclusions. On one hand, the strong functional dependence $q=q(\beta)$ documented across the full sample demonstrates that the system is operating near criticality in a regime where all 184 recordings, regardless of anatomical provenance, are parameterized by a single effective constraint. On the other hand, the spatial uniformity of $q$ across the STN boundary indicates that this near-critical nonextensive dynamical regime is not a local property of the STN alone but a feature shared by the entire basal ganglia neighborhood sampled intraoperatively. This observation is consistent with the view that Parkinson's disease  reorganizes the dynamical attractor of the cortico-basal-ganglia-thalamocortical loop, a network whose long-range correlations are constitutive of its normal operation, in such a way that dopamine depletion along the indirect pathway locks the system into a 
spatially extended, low-dimensional synchronized state, whose nonextensive statistical signature, as shown here, is uniform across anatomical boundaries \cite{chiken, parker, chen}.

\section{Final remarks}
The two main results of this work, the q-Gaussian goodness-of-fit across all recordings and the functional dependence $q \simeq 3-1.85\ \beta^{-0.33}$, place intraoperative Parkinsonian MER signals on firm theoretical ground within the framework of nonextensive statistical mechanics. Yet the theoretical significance of the present findings reaches considerably beyond the mere applicability of a generalized distribution. The collapse of 184 data points, sampled from 46 distinct patients and from two anatomically separate regions, onto a single monotonic curve in the $(q,\ 1/\beta)$ plane is the central result, and it demands a physical interpretation. In any empirical dataset, finding that two free parameters of a model family do not vary independently, but instead obey a robust functional law, means that the system is governed by fewer effective degrees of freedom than its formal description would suggest. In statistical mechanics, this reduction of dimensionality in parameter space is a hallmark signature of criticality: precisely at a critical point, the system's behavior is determined by a single relevant control parameter, and the susceptibility associated with the conjugate observable diverges. Identical phenomenology has been reported in the growth of asymptotically scale-free networks in various spatial dimensions \cite{BritoSilvaTsallis2016, NunesBritoSilvaTsallis2017, CinardiRapisardaTsallis2020} and in the acoustic emission preceding the fracture of construction materials, where the same power-law relationship between the $q$-Gaussian parameters was found to sharpen as the system approached its critical breakdown point \cite{GrecoTsallisRapisardaPluchinoFicheraContrafatto2020}. The present data demonstrate that living neural tissue in the parkinsonian state exhibits this same collective behavior.

The physical consequence of criticality that is most directly relevant to the neuroscience of Parkinson's disease is the divergence of susceptibility. In thermodynamic language, susceptibility measures how strongly a macroscopic observable responds to an infinitesimally small external perturbation of its conjugate field; at a phase transition, this response becomes unbounded, meaning that the system is in a state of maximal alertness to external inputs. In the neural context of the parkinsonian STN, this translates into a circuit that is structurally primed to amplify incoming signals out of proportion to their magnitude. The near-critical state identified in the MER amplitude statistics provides a quantitative, physics-grounded complement to the established electrophysiological description of Parkinsonism, linking the observed heavy-tailed fluctuations to the well-known beta-burst phenomenology of the parkinsonian basal ganglia network.

It is important, however, to distinguish the form of criticality reported here from the adaptive criticality often described in healthy cortical networks. In the healthy brain, near-critical dynamics are associated with high-dimensional attractors that support flexible and context-dependent motor behavior \cite{Shew}. In contrast, the near-critical state of the parkinsonian basal ganglia appears to be organized around a pathological, low-dimensional attractor sustained by dopamine depletion, in which susceptibility amplifies and stabilizes stereotyped beta-band oscillations.

From this perspective, therapeutic interventions such as deep brain 
stimulation can be interpreted as shifting the system away from this pathological 
low-dimensional attractor toward a dynamical regime of greater dimensionality and 
flexibility. In the language of $q$-statistics, what distinguishes healthy from pathological 
near-criticality is not the value of $q$ per se, which is expected to remain $q > 1$ in 
the healthy brain as a complex system with long-range correlations, but rather the nature 
of the near-critical constraint itself: the tight $q(\beta)$ coupling identified here is a 
hallmark of the parkinsonian attractor, and its disruption or relaxation, rather than any sensible 
shift of $q$ toward unity, constitutes the $q$-statistical signature of therapeutic action.

This interpretation provides a fresh and physically coherent perspective on the mechanism of DBS action. High-frequency stimulation of the STN has long been described as an "informational lesion" that overrides pathological rhythms, and it has been shown to reduce neuronal entropy in the parkinsonian primate model \cite{dorval}. Within the $q$-statistical framework, DBS can be understood as a perturbation that drives the system away from the near-critical regime identified here, disrupting the tight $q(\beta)$ coupling and, by extension, reorganizing the structure of the long-range correlations that sustain the heavy-tailed amplitude statistics, loosening the near-critical single-constraint without eliminating the nonadditive  character of the dynamics. The testable prediction that follows is that successful DBS might produce a weakening or dissolution of the tight $q(\beta)$ functional coupling, liberating the system's parameter space to recover more than one effective degree of freedom, while the value of $q$ itself is expected to remain $q > 1$, consistent with the intrinsic nonextensive nature of cortico-basal-ganglia dynamics. This would manifest as a dispersion of the $(q,\,\beta)$ data points away from the single-curve constraint of Eq.~(5), providing a distribution-geometry signature of therapeutic efficacy that is independent of and complementary to spectral beta-band power.

Beyond the STN and Parkinson's disease, the present results suggest a broader perspective on the use of $q$-statistics as a precision tool for characterizing the dynamical regime of the brain, in particular, for quantifying the proximity of a neural circuit to criticality and the structure of its long-range correlations, regardless of disease state. The $q>1$ values found uniformly across our sample indicate that the MER signals belong to the universality class of complex systems exhibiting nonlocal, power-law spatiotemporal correlations. These are precisely the conditions under which standard Boltzmann–Gibbs statistical mechanics loses its domain of validity and nonextensive statistical mechanics becomes the appropriate descriptive framework \cite{Tsallis1988, tsallis2004}. The fact that $q$ is spatially uniform across the STN boundary, with a ratio $\ \left\langle \bar{q}_{outside}/\bar{q}_{inside} \right\rangle =1.03$, further suggests that this statistical signature reflects a circuit-level, rather than a nucleus-specific property of the parkinsonian brain, consistent with the understanding that the pathological synchrony of PD involves the entire cortico-basal ganglia-thalamocortical loop as a dynamically coupled system. Analogous observations have been made in studies of epilepsy using $q$-Gaussians, where the degree of nonextensivity was found to increase progressively as the ictogenic network approached seizure onset, mirroring the transition from antipersistent to persistent dynamics observed in preseismic electromagnetic signals \cite{eftaxias} - a parallel that underscores the universality of the $q$-statistical framework across distinct forms of pathological criticality in complex physical and biological systems. 

\subsection{Limitations and Next Steps}

Several limitations of the present work deserve explicit acknowledgment. The analysis was conducted on a patient-balanced subsample drawn from a larger open-access dataset \cite{dataset}. Because recordings were sampled without clinical stratification, potentially relevant covariates such as disease duration, motor severity (UPDRS score), levodopa-equivalent medication dose at the time of surgery, and precise electrode depth within the STN remain uncontrolled. This design is appropriate for an exploratory analysis, but it leaves open the possibility that the parameters $q$ and $\beta$, or the functional dependence described in Eq.~(5), may vary systematically with these clinical factors.

A further methodological constraint concerns the scope of the statistical characterization performed here. The present analysis estimates only the stationary index $q_{stat}$ through $q$-Gaussian fitting of the MER amplitude distribution. The other two members of the $q$-triplet, $q_{sen}$, associated with the extensivity of $S_q$, and $q_{rel}$, describing the decay of temporal correlations, were not computed. Their estimation requires phase-space reconstruction and trajectory-divergence analysis for $q_{sen}$, and detailed autocorrelation analysis for $q_{rel}$, representing the most natural next step toward a full dynamical characterization of the parkinsonian basal ganglia.

The absence of healthy control recordings is another limitation that must be confronted. Because the dataset contains only MER recordings from patients with Parkinson's disease, it is not possible to determine whether the $q>1$ values and the near-critical $q(\beta)$ collapse are specific to the parkinsonian condition or instead represent generic features of basal ganglia MER signals. Addressing this question would require recordings from patients undergoing DBS for other movement disorders such as dystonia or essential tremor, or from trajectories incidentally traversing basal ganglia structures in non-PD contexts. Until such comparative data become available, the values of $q$ reported here should not be interpreted as exclusive markers of the parkinsonian condition. On theoretical grounds, $q > 1$ is expected for MER signals from any brain structure exhibiting long-range correlations, healthy or pathological; it is the specific form and tightness of the $q(\beta)$ constraint --- rather than the $q > 1$ character per se --- that this work puts forward as a candidate signature of near-criticality in the parkinsonian network. Future comparative recordings will be necessary to determine how the position along the $q(\beta)$ curve, and the strength of the coupling itself, differ between disease states.

From a spatial perspective, the binary inside/outside-STN labeling of the dataset precludes a depth-resolved analysis of how $q$ and $\beta$ evolve along the electrode trajectory. The STN contains well-defined functional subterritories—the dorsolateral sensorimotor zone, the ventromedial associative region, and the medial limbic susubdivision—and DBS literature consistently identifies the dorsolateral sensorimotor zone as the therapeutic ``sweet spot.'' Whether these subregions exhibit distinct $q$-statistical signatures, potentially providing an objective marker for optimal targeting, cannot be addressed with the present data and represents an important direction for future work. Similarly, extending the analysis along the full trajectory from the cortical entry point through the zona incerta, the fields of Forel, the STN, and into the substantia nigra could reveal whether a spatial $q$-profile might complement current qualitative approaches to STN localization.

Looking ahead, the most immediate clinical extension of the present framework is the real-time computation of $q$ from intraoperative MER signals. The grid-search fitting procedure used here is computationally lightweight, requires only short signal windows, and relies on minimal assumptions about stationarity beyond the estimation interval. These properties make it compatible with integration into existing surgical neurophysiology platforms, where a continuously updated $q$ profile could accompany electrode advancement, providing a distribution-level complement to beta-band spectral power and to the perceptual assessment of spike morphology currently performed by the surgical team. Such an implementation would also enable the construction of a spatial $q$ map of the DBS trajectory, an intraoperative atlas of nonextensivity that could be compared across patients and disease conditions.

In the longer term, applying the $q$-Gaussian framework to chronic local field potential recordings from implanted DBS systems would allow longitudinal tracking of $q$ and $\beta$ over extended periods of stimulation. Such data could test the hypothesis advanced in the preceding discussion: that effective DBS weakens the $q(\beta)$ coupling, disrupting the single-constraint parameterization that characterizes the near-critical parkinsonian attractor, without necessarily shifting $q$ toward unity, since nonextensive dynamics with $q > 1$ are expected to persist as an intrinsic property of cortico-basal-ganglia circuitry in both treated and untreated states. If confirmed, this relationship could provide a mechanistically grounded marker of therapeutic efficacy, potentially supporting future closed-loop DBS strategies based on real-time monitoring of the circuit's proximity to a pathological critical regime.


\bibliographystyle{unsrtnat}

\bibliography{cas-refs}

\end{document}